\documentclass[%
 reprint,
superscriptaddress,
 amsmath,amssymb,
 aps,
]{revtex4-2}

\usepackage{graphicx}
\usepackage{dcolumn}
\usepackage{bm}
\usepackage{listings}
\usepackage{fancyvrb}
\usepackage{xcolor}

\begin{document}

\renewcommand{\rq}{RydIQule}
\newcommand{\bra}[1]{$\langle #1 |$}
\newcommand{\ket}[1]{$| #1 \rangle$}

\preprint{APS/123-QED}

\title{\rq{}: A Graph-based \\Paradigm for Modelling Rydberg and Atomic Sensors}
%
%
\author{Benjamin N. Miller}
\affiliation{%
 Naval Air Warfare Center, 1 Administration Circle, China Lake CA 93555 USA
}%
\author{David H. Meyer}%
\affiliation{%
DEVCOM Army Research Laboratory, 2800 Powder Mill Rd, Adelphi, MD 20783 USA
}%
\author{Teemu Virtanen}%
\affiliation{%
 Naval Air Warfare Center, 1 Administration Circle, China Lake CA 93555 USA
}%
\author{Christopher M. O'Brien}%
\affiliation{%
 Naval Air Warfare Center, 1 Administration Circle, China Lake CA 93555 USA
}%
\author{Kevin C Cox}%
\affiliation{%
DEVCOM Army Research Laboratory, 2800 Powder Mill Rd, Adelphi, MD 20783 USA
}%

\date{\today}

\begin{abstract}
We describe a numerical technique and accompanying open-source Python software package called \rq{}. \rq{} uses a directional graph, relying on adjacency matrices and path-finding to generate a Hamiltonian for multi-level atomic systems.  \rq{} then constructs semi-classical equations of motion (Bloch equations) into a tensor which can store an entire simulation consisting of varied system parameters.  Using this framework, \rq{} returns solutions significantly faster than typical for interpreted programming languages. \rq{} extends beyond the capabilities of currently-available tools, facilitating rapid development in atomic and Rydberg spectroscopy. To demonstrate its utility, we use \rq{} to simulate a Doppler-broadened Rydberg atomic sensor that simultaneously demodulates five rf tones spanning from 1.7 to 116 GHz.  Using \rq{}, this simulation can be solved in several hours on a commercial off-the-shelf desktop computer.  
\end{abstract}

\maketitle

Atomic quantum sensors (e.g. clocks, magnetometers, electrometers, inertial sensors, etc.) are being used to solve real-world problems including  global positioning \cite{parkinson_navstar_1983}, imaging of biological systems \cite{xia_magnetoencephalography_2006}, and geodesy \cite{mcgrew_atomic_2018}, with new applications continually emerging.  The breadth of the atomic sensor design space is daunting, since one may utilize any combination of  atomic states, lasers, rf fields, time-dependence, atomic nonlinearities, laser-cooling and trapping, and Rydberg states.  Quantum sensor researchers need computationally-powerful software tools to serve as forward models for atomic vapor-based sensing.

One example of an emerging quantum sensor with  disruptive capabilities is the Rydberg atom-based electric field sensor \cite{mohapatra_coherent_2007, sedlacekMicrowaveElectrometryRydberg2012, andersonOpticalMeasurementsStrong2016, faconSensitiveElectrometerBased2016}.  Rydberg sensors can utilize dozens or more Rydberg states and detect time-dependent fields across the entire rf spectrum.  This structural richness allows for useful tuning schemes, THz imaging, simultaneous multi-band detection, and other potential use cases \cite{wade_real-time_2017, downes_full-field_2020,  fancher_rydberg_2021, simons_continuous_2021, meyer_simultaneous_2023}.    Further, these sensors often operate with room-temperature atoms, and modeling hundreds of velocity classes is necessary for accurate predictions.  For these reasons, modeling Rydberg sensors is computationally challenging over a large design space.  \rq{} and other complementary work can significantly aid the field, allowing rapid iteration and advances toward useful quantum sensing.

\rq{}, the subject of this work, complements previous research on  atomic physics numerical methods and solvers  \cite{johansson_qutip_2013, weber_calculation_2017, rochester_atomicdensitymatrix_nodate, eckel_pylcp_2020, patel_laser-atom_2022}.   The Alkali Rydberg Calculator (ARC) \cite{sibalicARCOpensourceLibrary2017}, for example, is now widely used to calculate matrix elements and energy levels of Rydberg states in Alkali atoms and is relied upon by \rq{}.  Although mathematical graphs have recently been used for atomistic calculations of materials \cite{choudhary_atomistic_2021}, we are not aware of their use for atomic quantum solvers.  \rq{} complements and extends current libraries by providing computationally efficient generation of quantum equations of motion (Bloch equations) as well as solvers that can include many Rydberg states, complex and closed-loop level diagrams, rf fields, time dependence, and Doppler-broadening. \rq{} solves many such systems in seconds or minutes.

In this Article, we first describe \rq{}'s graph architecture and show how it is used to generate Hamiltonians and equations of motion for multi-level and Rydberg atomic systems.  We also discuss how multi-parameter systems are stacked into tensor equations for rapid solving.  We present a minimal (9-line) pseudo-code example of how \rq{} is used.  We show that \rq{}'s architecture enables a significant advance in speed for forward modeling of atomic and Rydberg quantum sensors.  Next, we demonstrate a \rq{}-based simulation of a recent Rydberg atomic sensing experiment \cite{meyer_simultaneous_2023}, including five Rydberg states and five time-dependent rf signals.  This scheme uses Rydberg heterodyne detection to simultaneously receive the amplitude and phase of five tones ranging from 1.7 to 116 GHz, a task that would likely be difficult with a classical receiver.  The simulation utilizes approximately eight million solver calls and  solves in a few hours on a commercial off-the-shelf desktop.  The full source code for this simulation is included as supplemental material \cite{cox_supplemental}. The software itself, including a full user manual and documentation, can be found at https://github.com/QTC-UMD/rydiqule.

\begin{figure}
    \centering
    \includegraphics[width=0.5\textwidth]{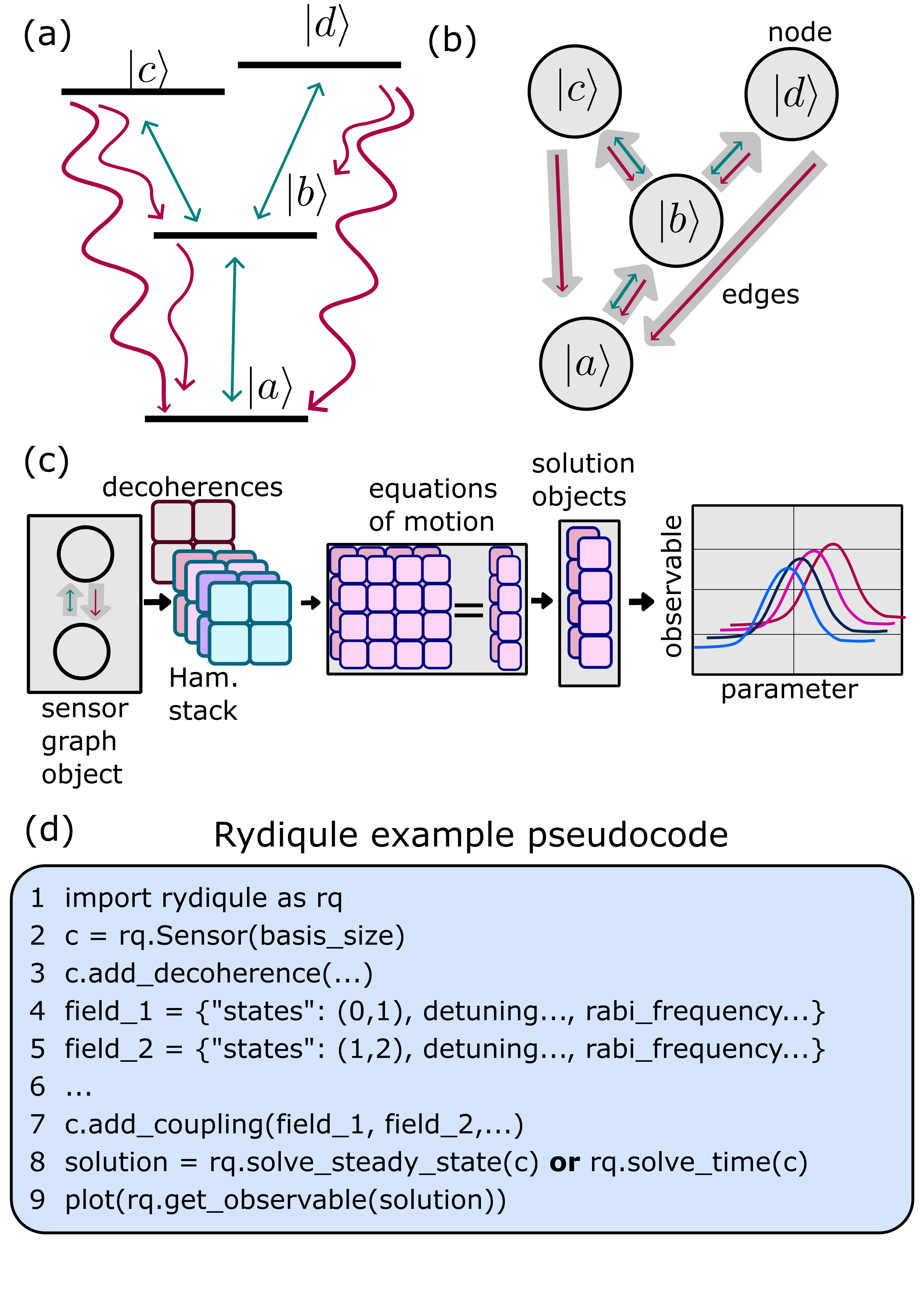}
    \caption{\rq{} graph representation of an atomic system.  (a) example level diagram with its corresponding graph representation in (b). (c) flow diagram of a \rq{} simulation. (d) Pseudo-code to perform a simulation with \rq{}. }
    \label{fig:graph}
\end{figure}

We first discuss the key elements of \rq{} distinguishing it from previous work that make it uniquely powerful for modeling atomic and Rydberg systems, shown in Figure \ref{fig:graph}(a-c). The atomic energy diagram (a) is stored as a directed graph using the NetworkX python package \cite{SciPyProceedings_11}. Atomic levels and their properties are stored as nodes, and field couplings (either laser or rf) between them are stored as edges.  NetworkX stores graph nodes and edges as Python dictionaries, allowing multiple edge weights and node attributes. This accommodates a wide range of parameters, including transition frequencies, detunings, laser powers, and decoherence rates, to be stored on the graph. A single graph object of this sort is wrapped in \rq{}'s central class, called a sensor, and contains all information about an atomic sensor model, including multiple swept experimental parameters that are defined as arrays. Once a sensor is defined with levels, couplings, and parameters, \rq{} can compute its Hamiltonian matrices, corresponding equations of motion, solutions to those equations, and associated physical observables. 

\rq{}'s graph-based representation of the atomic system provides a number of key benefits. The first and most obvious of these is flexibility. The graph architecture is general enough to store level diagrams with arbitrary connectivity and provides an intuitive visualization of the level diagram.
Second, the graph is used to construct the Hamiltonian $H$.  The off-diagonal elements $H_\text{OD}$ and the diagonal $H_\text{D}$ are computed separately and summed together $H = H_\text{D} + H_\text{OD}$.  $H_\text{OD}$ is equal, by construction, to an adjacency matrix of the graph, with edge weights given by the Rabi frequency of each coupling.  The decoherence matrix, used to compute the Langevin terms of the master equation, can also be computed from an adjacency matrix, with weights equal to the decoherence rates.
In the rotating wave approximation, the $i$th diagonal term of $H_\text{D}$ is, by definition, given by the path ``distance'' $x_i = \sum_{j \in P_i} \delta_j$ from each node $i$ to the ground state node (defined at zero distance on the graph),
where the distance of each edge (coupling field) is given by its detuning $\delta_j$, and $j$ indexes the edges along path $P_i$.  We use Dijkstra's path-finding algorithm, built into NetworkX, to find each path $P_i$ \cite{dijkstra1959note, SciPyProceedings_11}.  Couplings can also be defined by their absolute frequency with no rotating wave approximation is applied.  

These algorithms applied to the graph structure give \rq{} flexibility to correctly construct Hamiltonians for ladder, $\Lambda$, or V-schemes, systems with dozens or more states, hyperfine sublevels, dark states, and time-varying couplings.  Diamond schemes that include full loops of field couplings cannot be self-consistently represented in the rotating frame.  However, these situations can still be treated easily in \rq{} by specifying one of the loop couplings as time dependent. A full description of the Hamiltonian generation can be found in the documentation and source code \cite{noauthor_rydiqule_2023}.

\rq{} relies on the ease-of-use features of the python language, making it simple to define a system. However, when modeling physical systems using popular interpreted languages such as Python, Matlab, or Mathematica, the time-cost of code interpretation can easily exceed the compute time of a bespoke, compiled solver by orders of magnitude. This often leads to the approach of maximizing computing hardware and/or waiting extended periods of time for results.  Innovations occur faster when computations can be rapidly iterated on consumer-grade desktops and laptops.  Balancing these competing concerns has been core to our design.

\rq{} mitigates code interpretation slowdowns by making use of compiled NumPy routines \cite{harris_array_2020} for challenging calculations that scale with the problem size.  NumPy's routines are written in C, and are the industry standard for efficient matrix calculations. \rq{}'s  goal is to contain its computational complexity into these functions. The primary way that our framework accomplishes this is via a technique \rq{} calls ``stacking", which is core to the design of many functions within NumPy.  For basis size (i.e. number of energy levels) $b$, \rq{} constructs Lindblad master equations of motion (EOM) of size $n\times n$ in the superoperator form where $n = b^2-1$ \cite{manzano_short_2020}.  \rq{} asserts population conservation to eliminate one of the $b^2$ equations.  If a Hamiltonian parameter (e.g. coupling detuning or power) is scanned over $d_1$ values, rather than treating $d_1$ individual $n\times n$ matrices, \rq{}  arranges the EOM into a single $d_1 \times n \times n$ NumPy array (that is, a rank three tensor). Similarly, $d_2$ such arrays can be arranged into a single $d_2 \times d_1 \times n \times n$ array, and so on. We refer to the ``stack shape" $^*l$ as the array shape of the parameter dimensions (i.e. [$d_1, ..., d_p$]) for $p$ independent parameters. This treatment allows for parameter spaces of arbitrary dimension to be represented as a single NumPy array. Rather than performing calculations individually in Python and introducing interpreted-language overhead, matrices are built into a single object that can be manipulated with a single call to a compiled NumPy function.

A minimal pseudo-code example of \rq{} is shown in Fig. \ref{fig:graph}(d).  A simulation is performed using 5 steps.  First, after \rq{} is imported, a Sensor object is created using the Sensor class, to store all of the information about the simulation (line two).  Next, the decoherence rates between atomic levels are defined (line 3).  The laser, microwave, and rf field couplings are defined and added (lines 4-6).  Finally, we call a solver function, passing the sensor object as a parameter (line 8).  A number of functions exist to retrieve and plot the observables that would be obtained by a real experiment (line 9).

To roughly quantify the performance of \rq{}, we first consider the theoretical expectations of its time complexity. We consider a single set of equations of motion to be of size $n \times n$.  We allow $p$ independent parameters, where the $i^{th}$ parameter is scanned over an array of values with length $d_i$.  To find a steady-state solution, this matrix is diagonalized by Gaussian elimination, which is known to have time cost proportional to the number of rows $n$ cubed.  Thus, we expect a total solve time $t \propto  b^5 \prod_{i=1}^p d_i \propto n^3 \prod_{i=1}^p d_i$.   In the case where each parameter's dimension is the same length $d$, the product term above simplifies to an exponential $d^p$.

\begin{figure}[t]
    \centering
    \includegraphics[width=0.5\textwidth]{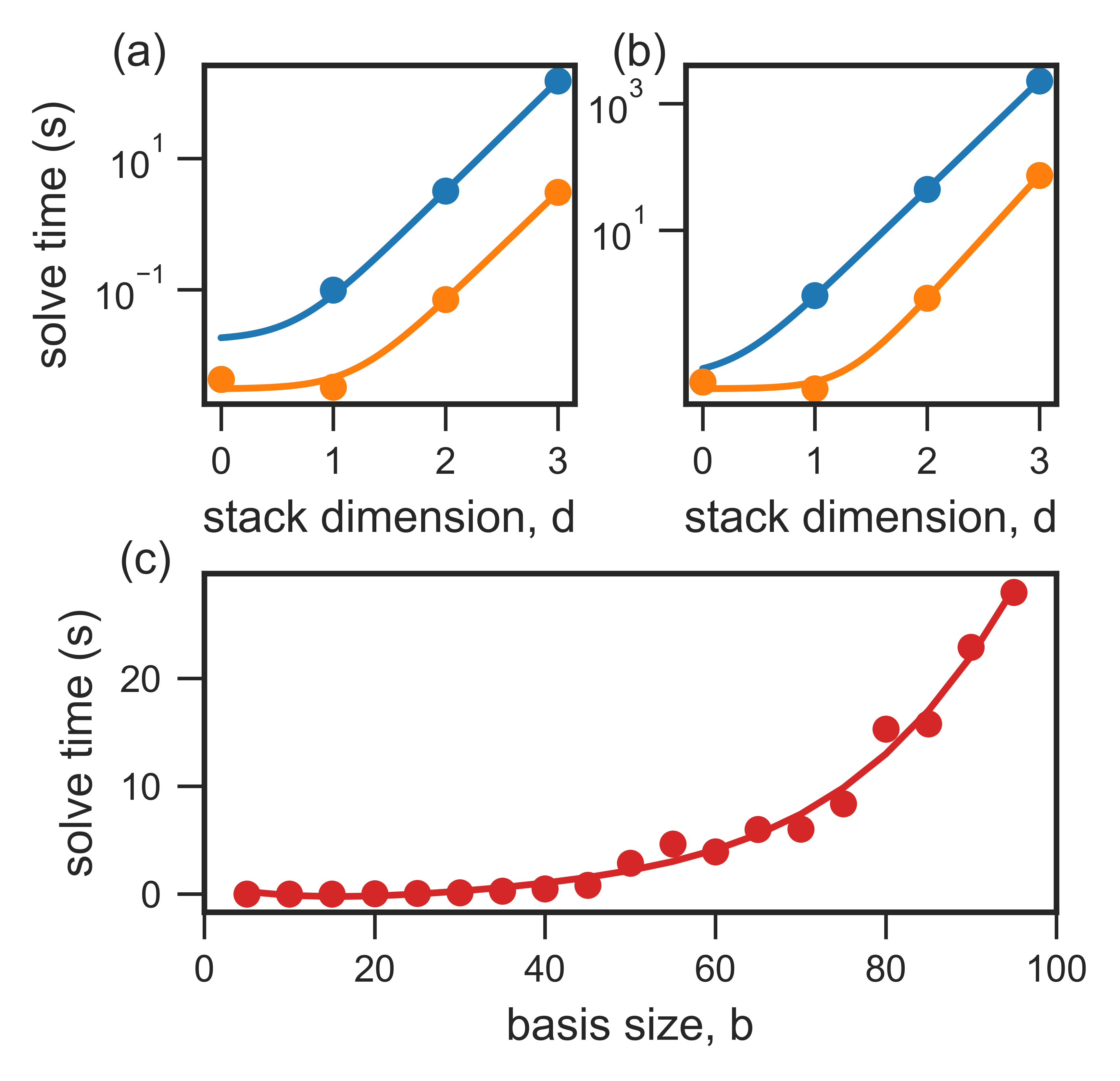}
    \caption{Time to solve a ladder scheme in \rq{}. (a-b) Comparison of python loop (blue) vs \rq{} ``stacked" (orange) solve time in parameter sweeps over varying numbers of parameters, referred to as stack dimension, in the (a) steady-state and (b) time domain, and exponential fits. (c) Time to solve a steady-state system versus the number of quantum states in the system, along with the best 5th-order polynomial fit.}
    \label{fig:rq-speed}
\end{figure}

We confirm the two scalings, $t \propto b^5$ and $t \propto d^p$, with results shown in Fig \ref{fig:rq-speed}.  We also show that \rq{}'s stacking method solves faster than explicitly looping over parameters in Python. For (a) and (b), we define 4-level systems with three couplings in a ladder configuration, treating all couplings in the rotating wave approximation. We then solve the system in the steady state (a) and time domain for 50 microseconds (b) and measure the time for 0, 1, 2, and 3 of the couplings scanned over a range of 25 detuning values. The fit curve is an exponential function plus a constant to account for roughly constant overhead.  Both stacking and looping solve in time proportional to $d^p$ but interpreted python loops lead to an additional slowdown of order $\times 100$. In Fig. \ref{fig:rq-speed}(c), we time \rq{}'s steady-state solver versus basis size. The data is fit to a 5th order polynomial, indicating the expected scaling. 
 The IPython notebooks used to generate this data are included as Supplemental Material \cite{cox_supplemental}.
 
Stacking is computationally efficient for code written in Python since the entire simulation is solved using pre-compiled NumPy routines. However, this requires that the entire model fits into the computer's memory.  This condition is often broken, especially when considering room-temperature atoms with Doppler averaging.  For this reason, \rq{} can ``slice'' the stack into chunks that fit in the computer's memory, minimizing the number of calls necessary to the Python interpreter. For example, if a given system of equations would require four times the computer's available memory to solve, \rq{} will automatically break the equations into four pieces and solve them individually. It is with this combination of stacking and slicing that \rq{} handles large simulations efficiently with no additional effort on the part of the user.

\begin{figure}
    \centering
    \includegraphics[width=0.5\textwidth]{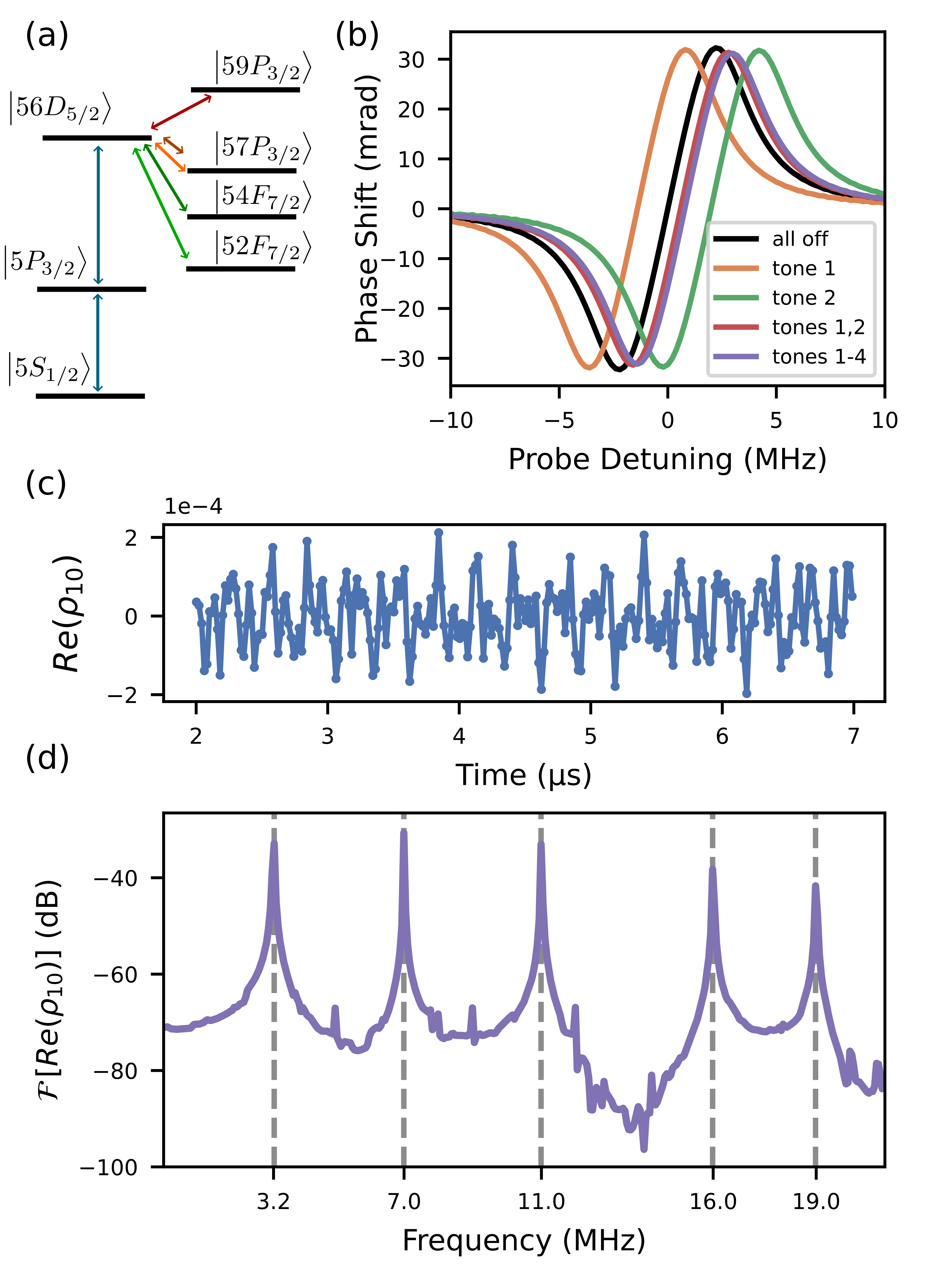}
    \caption{\rq{} model for multitone reception based on Ref. \onlinecite{meyer_simultaneous_2023}. (a) Level diagram showing the five Rydberg states and five rf tones.  (b) Simulated phase shift of the probing laser versus probe laser detuning for several combinations of the rf LO fields.  (c) Normalized time response of the density matrix for a fixed probe detuning with the five RF tones applied to the simulated Rb85 cell. (d) (i) Fourier Transform of the data shown in figure \ref{fig:multitone} (b). We recover the signal frequencies, as indicated by the vertical dashed lines.}
    \label{fig:multitone}
\end{figure}

In the final part of this Article, we simulate a recent experiment that requires the full set of \rq{}'s capabilities.   The experiment demodulated five rf fields simultaneously using two lasers and four Rydberg transitions \cite{meyer_simultaneous_2023}.  The source code (Python Jupyter notebook) to create this data is included in the Supplemental Material \cite{cox_supplemental}, and can be modified to simulate a large variety of similar schemes.  As constituted, this time-dependent simulation took around one minute to complete on a modern 16 core desktop with 256 GB of memory, and 4-6 hours on the same machine including atomic motion.  

The level diagram is shown in Fig. \ref{fig:multitone}(a).  The system uses the Rydberg rf heterodyne technique, which relies on the off-resonance square-law response of the atoms to detect the amplitude and phase of an incoming carrier \cite{jauVaporCellBasedAtomicElectrometry2020, meyerWaveguideCoupledRydbergSpectrum2021,hu_continuously_2022}.  For simulation purposes, each rf tone is represented as $E^i(t) = E^i_{LO}\sin(\omega^i_0 t) + E^i_{sig}(t) \sin(\omega^i_0 t + \omega^i_m t)$, where $\omega^i_0$ is the local oscillator detuning for the $i$th tone, $\omega^i_m$ is the baseband signal frequency, $E^i_{LO}$ is the strength of the local oscillator (LO) portion of the field, and $E^i_{sig}$ is the strength of the signal field.

Figure \ref{fig:multitone}(b) shows the spectroscopy traces that result from scanning the detuning of the probe laser when each tone is applied.  The signal is plotted in terms of optical phase, that would be retrieved from a homodyne detection system.  For convenience, the local oscillators are arranged so that the total AC Stark shift nominally cancels.  The real component of the probing density matrix element $\rho_{10}$, that is proportional to the sensor's output voltage  $V \propto Re(\rho_{10}) \propto \sum_i E^i_{sig} \sin(\omega^i_m t)$, is displayed in Fig. \ref{fig:multitone}(c).  This time-domain signal includes beats from all applied tones that can be deciphered through its Fourier transform denoted by $\mathcal{F}[\,]$ (Fig \ref{fig:multitone}(d)).  The Fourier data shows each heterodyne beat.   In the \rq{} documentation \cite{noauthor_rydiqule_2023}, we demonstrate additional tools to analyze the signal-to-noise ratio and noise-equivalent field sensitivity to incoming fields. 

\rq{} significantly advances publicly-available capabilities to simulate atomic sensors, but further work in \rq{} and other supporting projects is still needed.  \rq{} is currently semi-classical.  Electromagnetic fields are treated as complex parameters in each equation, with no atomic back action, meaning that \rq{} is most accurate when the optical depth is low \cite{meyer_optimal_2021}. Furthermore, \rq{} is a single-atom solver that does not explicitly handle atom-atom
interactions or quantum entanglement between atoms.  These approximations are often valid for room-temperature thermal vapor-based atomic spectroscopy in free space, in applications such as atomic magnetometers and in particular, Rydberg electrometers.  But further development will be required to access non-classical applications involving atom-atom and atom-light entanglement.

One of the most significant computational overheads in \rq{} arises from handling thermal vapors.  Doppler broadened ensembles require more stacks since we must solve the equations for all velocity classes. Due to detailed structure in the atomic frequency response, \rq{} typically stacks hundreds of velocity classes for accurate results.  Significant additional speedups are likely possible using more efficient pre-compilation of the time integration, graphical processing units, or methods to improve the efficiency of sampling atomic motions \cite{rotunno_inverse_2023}. 

We hope that this manuscript will inspire scientists to modify, change, and recreate \rq{} and similar packages for a wide range of applied physics applications.  The key advances, including the graph structure and overall python framework, can be extended to a wide range of applications including other quantum sensors, quantum memories, and multi-level spectroscopy.  Open-source tools to validate models of atomic devices will be a catalyst for the development and application of quantum science. 

\begin{acknowledgments}
The authors acknowledge helpful advice, discussions, and contributions from Paul Kunz,  Joshua Hill, Peter Elgee, Fredrik Fatemi, Nelson Li, and William Wolfs.  The authors acknowledge funding from the Defence Advanced Research Programming Agency (DARPA).  Benjamin Miller, Christopher O’Brien, and Teemu Virtanen recognize financial support from the Office of Naval Research (ONR) In-House Laboratory Independent Research (ILIR) program at the Naval Air Warfare Center Weapons Division. The views, opinions and/or findings expressed are those of the authors and should not be interpreted as representing the official views or policies of the Department of Defense or the U.S. Government.
\end{acknowledgments}

\bibliography{bib1}

\end{document}